\documentclass{emulateapj}
\usepackage{apjfonts}
\usepackage{epsfig}
\usepackage{natbib}

\newcommand{\chisq}{\ensuremath{\chi^2}}

\newcommand{\chandra}{\emph{Chandra}}

\def\gtrsim{\mathrel{\hbox{\rlap{\hbox{\lower4pt\hbox{$\sim$}}}\hbox{\raise2pt\hbox{$>$}}}}}

\newcommand{\fwoiii}{\ensuremath{\mathrm{FWHM}_\mathrm{[O \tiny III]}}}

\newcommand{\hbeta}{H\ensuremath{\beta}}

\newcommand{\hst}{\emph{HST}}

\newcommand{\kms}{km~s\ensuremath{^{-1}}}

\newcommand{\lbol}{\ensuremath{L_{\mathrm{bol}}}}
\newcommand{\loiii}{\ensuremath{L_{\mathrm{[O {\tiny III}]}}}}

\newcommand{\msun}{\ensuremath{M_{\odot}}}

\newcommand{\oii}{[\ion{O}{2}]}
\newcommand{\oiii}{[\ion{O}{3}]}

\newcommand{\sii}{[\ion{S}{2}]}

\newcommand{\xmm}{{\it XMM-Newton}}

\newcommand{\bub}{SDSS~J1356$+$1026}

\def\lax{{$\mathrel{\hbox{\rlap{\hbox{\lower4pt\hbox{$\sim$}}}\hbox{$<$}}}$}}
\def\gax{{$\mathrel{\hbox{\rlap{\hbox{\lower4pt\hbox{$\sim$}}}\hbox{$>$}}}$}}

\slugcomment{Dec 14, 2011;
to be published in {\it The Astrophysical Journal}.}
\shorttitle{{\it SDSS J1356+1026}}
\shortauthors{GREENE, ET AL.}

\begin{document}

\title{A spectacular outflow in an obscured quasar}

\author{Jenny E. Greene}
\affil{Department of Astrophysical Sciences, Princeton University, Princeton, 
NJ 08544}

\author{Nadia L. Zakamska}
\affil{Kavli Institute for Particle Astrophysics and Cosmology,
  Stanford University, 2575 Sand Hill Road, MS-29, Menlo Park, CA
  94025; Kavli Fellow\\
Department of Physics and Astronomy, Johns Hopkins University,
Bloomberg Center, 3400
N. Charles St., Baltimore MD 21218}

\author{Paul S. Smith}
\affil{Steward Observatory, University of Arizona, Tucson, AZ 85721}

\begin{abstract}
  \bub\ is a pair of interacting galaxies at redshift $z=0.123$ that
  hosts a luminous obscured quasar in its northern nucleus.  Here we
  present two long-slit Magellan LDSS-3 spectra that reveal a pair of
  symmetric $\sim$ 10 kpc-size outflows emerging from this nucleus,
  with observed expansion velocities of $\sim$250 \kms\ in
  projection. We present a kinematic model of these outflows and argue
  that the deprojected physical velocities of expansion are likely
  $\sim 1000$~\kms\ and that the kinetic energy of the expanding
  shells is likely $10^{44-45}$ erg~s$^{-1}$, with an absolute minimum
  of $>10^{42}$ erg~s$^{-1}$. Although a radio counterpart is detected
  at 1.4GHz, it is faint enough that the quasar is considered to be
  radio-quiet by all standard criteria, and there is no evidence of
  extended emission due to radio lobes, whether aged or continuously
  powered by an on-going jet. We argue that the on-going star
  formation is probably insufficient to power the observed energetic
  outflow and that \bub\ is a good case for radio-quiet quasar
  feedback.  In further support of this hypothesis, polarimetric
  observations show that the direction of quasar illumination is
  coincident with the direction of the outflow.
\end{abstract}

\section{Introduction}

The discovery of a tight relationship between the masses of black
holes (BHs) in nearby galaxies and the velocity dispersions and masses of their
stellar populations \citep{magorrianetal1998, gebhardtetal2000a,
  ferraresemerritt2000} suggests that the active phase of the
supermassive BH evolution has profound effects on the
formation of the galaxy. Indeed, there is direct evidence for such
`feedback' by radio-loud active galactic nuclei (AGNs) onto the
large-scale gaseous environment of their host galaxies, both at low
and at high redshifts \citep{nesvadbaetal2008, fustockton2009}. In
these objects, ionized gas is outflowing with velocities above the
escape speed from the galaxy, often along the direction of the radio
jet. This also appears to be the mode in which AGNs heat up galaxy
clusters \citep{mcnamaranulsen2007}. However, as only a minority
($\sim$10\%) of active BHs are radio-loud, this mechanism may not be
sufficient to explain the colors and the luminosity function of
massive red galaxies (e.g., \citealt{springeletal2005}), 
and it is as yet unclear if the remaining $\sim$90\% of AGNs are affecting the gas, 
and thus the star formation, in the host galaxies in a similar fashion.  

There is strong evidence that star formation driven winds 
play an important role in galaxy evolution.  Their imprint is apparent 
in the mass-metallicity relation at low redshift \citep[e.g.,][]{tremontietal2004}, 
and they are required in galaxy simulations to reproduce the 
observations \citep[e.g.,][]{springeletal2005,daveetal2008}.  Most importantly, 
winds driven by star formation are clearly seen in low-redshift dwarf galaxies 
\citep[e.g.,][]{martin1998}, in local ultra-luminous infrared galaxies 
\citep[ULIRGs; e.g.,][]{rupkeetal2005}, in galaxies at $z\sim1$ 
\citep[e.g.,][]{weineretal2009} and high-redshift star-forming galaxies as well 
\citep[e.g.,][]{shapleyetal2003}.  See \citet{veilleuxetal2005} for a recent review of 
low redshift observations.  What has been much harder to pin down is the role of 
radio-quiet BH accretion in driving winds 
\citep[e.g.,][]{tremontietal2007,alexanderetal2010,krugetal2010,
coiletal2011,hainlineetal2011,rupkeveilleux2011}.

One class of BHs that may be showing radio-quiet feedback processes in
action are broad absorption line quasars, in which high-velocity
outflows appear to be driven by the radiation of the quasar itself
\citep[e.g.,][]{progaetal2000}. Because the distance of the outflowing
material from the BH and its covering factor are difficult to
determine, the kinetic energies of these outflows are unknown in most
cases.  In some sources, it may be a substantial fraction of the
accretion energy \citep{crenshawetal2003, moeetal2009}, and the
detailed photo-ionization models place these outflows at a
considerable distance (several kpc) from the center of the galaxy
\citep[but see also][]{faucher-giguereetal2011}.  These observations
are promising, but so far have been possible only for a handful of
sources.  We seek unambiguous evidence for galaxy scale winds.

A handful of strong cases have been made for radiatively driven AGN
outflows on galaxy-wide scales at high redshift
\citep[e.g.,][]{tremontietal2007,alexanderetal2010,hainlineetal2011}.  Locally, new
exciting observations of molecular lines, both with ground-based
interferometers \citep[][]{alataloetal2011} and with Herschel
\citep[e.g.,][]{fischeretal2010} are revealing large molecular
outflows that are plausibly AGN-driven.  In particular,
\citet{sturmetal2011} make a strong case that the molecular outflow
velocity in local ultra-luminous infrared galaxies (ULIRGs) is
strongly correlated with the strength of the active nucleus, such that
more luminous AGNs drive faster outflows \citep[although see
also][]{krugetal2010}.

Perhaps the best low-redshift case study of large-scale outflows is of
Mrk 231.  Mrk 231 is the most luminous local ULIRG
\citep{sandersetal1988}, the nearest broad emission-line AGN, a
rare low-ionization broad absorption-line quasar, and contains a radio
jet.  This galaxy also hosts a massive young starburst that likely
dominates the bolometric output of the galaxy
\citep[e.g.,][]{lonsdaleetal2003}.  The galaxy has been studied in
detail at every conceivable wavelength \citep[see, e.g., ][and
references therein]{feruglioetal2010}.  Most relevant to our work are
the recent results pointing to a massive molecular
\citep{fischeretal2010,feruglioetal2010} and neutral
\citep{rupkeveilleux2011} outflow in this galaxy.  It is argued
strongly in these papers that radiative driving, rather than star
formation or jet activity, power the observed outflows.  In this paper
we try to make a similar case for radiation-pressure driving, except
that we focus on a radio-quiet obscured quasar, for which there is
little evidence of massive star formation in the optical
spectrum.  By focusing on obscured, radio-quiet systems, we hope to
isolate unambiguously the sources of acceleration in the observed
outflows.  We also hope to test the hypothesis that AGN feedback
occurs most vigorously in the obscured phase
\citep[e.g.,][]{sandersetal1988,hopkinsetal2006}.

Since 2007, we have been conducting a study of the morphologies and
kinematics of narrow-line regions in obscured radio-quiet quasars
\citep{zakamskaetal2003, reyesetal2008}, in search of tell-tale signs
of the interactions between the supermassive BH in its radiative
(rather than jet-launching) phase and its large-scale gaseous
environment. In our program conducted with the Magellan telescopes
(\citealt{greeneetal2009}, hereafter Paper I, and
\citealt{greeneetal2011}, hereafter Paper II), we obtained 15
long-slit spectra of obscured quasars. We find that in the quasar
luminosity regime probed by our observations (\loiii$\approx
10^{42}-10^{43}$~ erg~s$^{-1}$ or inferred \lbol$\approx
10^{45}-10^{46}$~ erg~s$^{-1}$), the quasar easily ionizes the galaxy
along any direction not obscured by dust.  Furthermore, the gas is
disturbed on galaxy-wide scales, with both broad emission-line widths and
kinematics clearly inconsistent with galaxy rotation. Among these 15
sources, one object, SDSS J135646.10+102609.0 (hereafter SDSS J1356+1026), 
shows very extended ($\ga 50$ kpc) ionized line
emission which is highly kinematically organized on $\sim 20$ kpc
scales. In this paper, we focus on this source, and build
a case for AGN-driven feedback using an analysis of the
kinematics and geometry of the ionized gas.

\section{Observations and Basic Object Properties}

Despite its anonymous telephone number, the enigmatic object
\bub\ has already
garnered quite a bit of interest in the recent literature. It was
targeted for spectroscopy by the Sloan Digital Sky Survey
\citep{yorketal2000} in the Data Release 4 \citep{adelmanetal2006}. Due to its
large \oiii$\lambda 5007$\AA\ (hereafter \oiii) luminosity and
line-ratios indicative of an active nucleus, it was included in the
catalog of obscured (type 2) quasars published by \citet{reyesetal2008}. The
kinematic structure of the \oiii\ line (as seen within the 3\arcsec\
SDSS spectroscopic fiber) is complex, showing at least two
components. As a result, \citet{liuetal2010survey} included \bub\ in a
sample of type 2 AGNs with two double-peaked emission lines. 

\begin{figure*}
\vbox{ 
\vskip -0.truein
\hskip 0.75in
\psfig{file=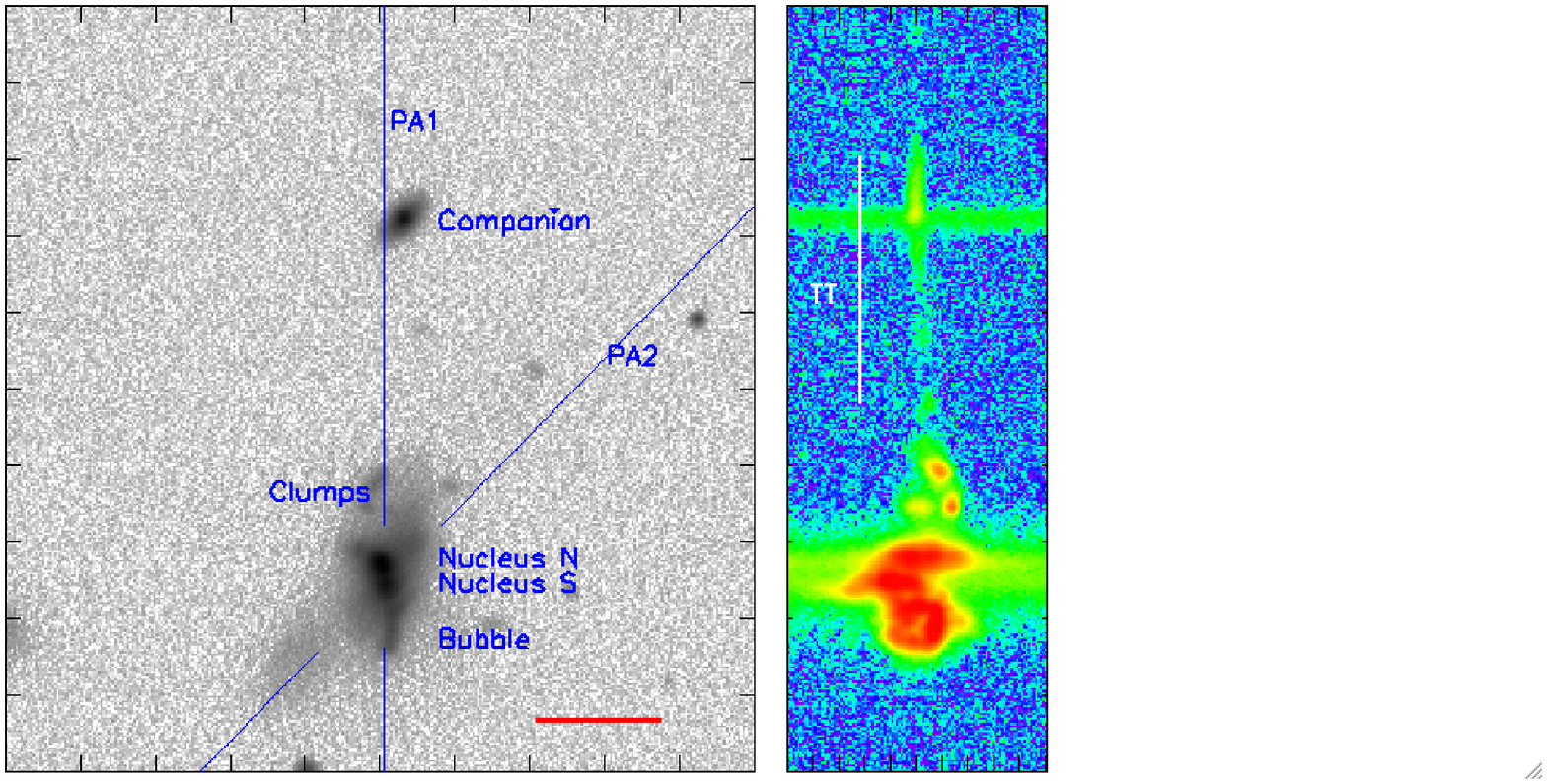,width=0.95\textwidth,keepaspectratio=true,angle=0}
}
\figcaption{
{\it Left}: $r-$band Magellan acquisition image of \bub\ and its immediate
environment. Red scale bar indicates 10\arcsec\ (22 kpc) and blue
lines indicate positions of the two spectroscopic slits. N is up and E
is to the left. The observation taken with the N-S slit is PA1 and the
one with the SE-NW slit is PA2.  
{\it Right}: The two-dimensional
spectrum centered on [O {\tiny III}]$\lambda 5007$ taken at the N-S
slit position.  The vertical scale is the same as that of the image on
the left. The full horizontal span is 3842 \kms\ centered at 5010\AA\
(rest frame).  The ``bubble'' feature that is the focus of this paper
is the round feature to the south of the continuum sources.  There is
also a tidal feature (TT) connecting the merging galaxies with the
companion to the north, with the extent demarcated by the vertical
white line.
\label{fig:bubbleim}}
\end{figure*}

\vbox{ 
\vskip +45mm
\hskip +0mm
\psfig{file=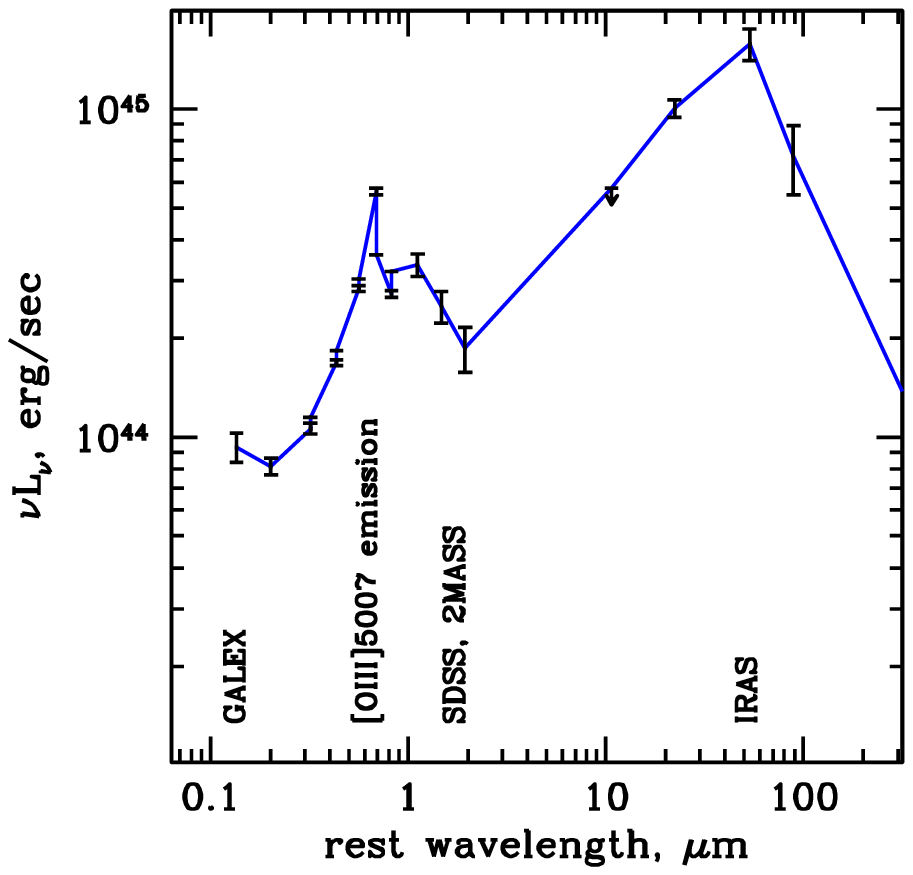,width=0.4\textwidth,keepaspectratio=true}
}
\vskip -0mm
\figcaption[]{The broad-band spectral energy distribution of \bub.  Error bars represent the 
data presented in Table 1, while the blue line simply connects the points to guide 
the eye.
\label{fig:SED}}
\vskip 5mm
\noindent
Such signatures can be due either to the kinematic structure of the
narrow-line region itself or to two quasars, each with its own
narrow-line region, with a velocity offset. The galaxy morphology is
clearly disturbed, suggesting an on-going merger (Figure \ref{fig:bubbleim} {\it
  left}). The presence of two merging galactic nuclei in \bub\ is
apparent both in optical
\citep{greeneetal2009,greeneetal2011,fuetal2011ao} and in near-infrared
\citep{shenetal2011} observations, and strong high-ionization emission
lines are associated with both (Paper II and this paper).

The spectroscopic observations of \bub\ were obtained on UT March 18,
2007 using the Low-Dispersion Survey Spectrograph \citep[LDSS3;
][]{allington-smithetal1994} with a $1\arcsec \times 4\arcmin$ slit at
the Magellan/Clay telescope on Las Campanas.  The seeing was $\sim
1\arcsec$. We observed with two slit positions for one half hour at
each position.  Slit position 1 (PA1 hereafter) was oriented
north-south, while slit position 2 (PA2 hereafter) was oriented at $45
\degr$ to the west of PA1 (Figure \ref{fig:bubbleim}) . We used the
VPH-Blue grism in the reddest setting, resulting in wavelength
coverage of $4300-7050$ \AA\ and velocity resolution of $\sigma_{\rm
  inst}\approx 67$~\kms. In addition to the primary science targets,
at least two flux calibrator stars and a number of velocity template
stars were also observed.

Cosmic-ray removal was performed using the spectroscopic version of
LACosmic \citep{vandokkum2001}. Bias subtraction, flat-field
correction, wavelength calibration, pattern-noise removal (see Paper
I), and rectification were performed using the Carnegie Observatories
reduction package
COSMOS\footnote{http://obs.carnegiescience.edu/Code/cosmos}. 
The flux calibration correction is determined from the
extracted standard star using IDL routines following methods described
in \citet{mathesonetal2008} and then applied in two dimensions.  In
Paper I we demonstrate that our flux calibration is reliable
at the $\sim 40\%$ level.

Polarimetric observations were made at the 1.54m Kuiper telescope with
SPOL spectropolarimeter \citep{schmidtetal1992a} on March 29, 2011,
with the 3''-wide slit oriented in the E-W direction and an
integration time of 32 minutes. 
The same instrument was again used at the 2.3m Bok telescope on May
27, 2011 with the 3"-wide slit oriented in the N-S direction and an
integration time of 40 minutes.  
In both cases, the polarization
measurements span 4000-7550\AA\ with a resolution of $\sim 20$
\AA. The degree of linear polarization (P) is corrected for statistical bias
\citep{wardlekronberg1974} and the position angle of the polarization
($\theta$) was calibrated by observing the interstellar polarization
standard stars Hiltner 960 and VI Cyg \#12 \citep{schmidtetal1992b}.
Further details of the general observation and reduction procedures
for the spectropolarimetry can be found in \citet{smithetal2003}.

We have compiled the spectral energy distribution of \bub\ based on
publicly available catalogs, from radio to ultra-violet wavelengths
(see Table 1; Figure \ref{fig:SED}).  X-ray observations with both
\xmm\ and \chandra\ are currently underway.

\section{Analysis of observations}

\subsection{Observed emission}
\label{sec:obsem}

We first introduce the basic observational components of \bub\ and its
immediate environment, as seen in Figure \ref{fig:bubbleim}.  We
observe two nuclei, the northern (N) nucleus and the southern (S)
nucleus, separated by $\sim 2.5$ kpc (1.1\arcsec) on the
sky. We find distinct stellar continuum and ionized gas emission
from each nucleus in \bub.  Although the continua are spatially
resolved, the southern source is faint enough that we are unable to
determine the velocity separation between the nuclei from the stellar 
spectra.  To the north is a companion galaxy located 62 kpc (in
projection on the plane of the sky) from the N nucleus. In velocity
space, the stellar features of the companion galaxy are blueshifted by
$\sim 250$ \kms\ with respect to the stellar features of the N
nucleus. 

\begin{figure*}
\vbox{ 
\vskip -0.truein
\hskip 0.5in
\psfig{file=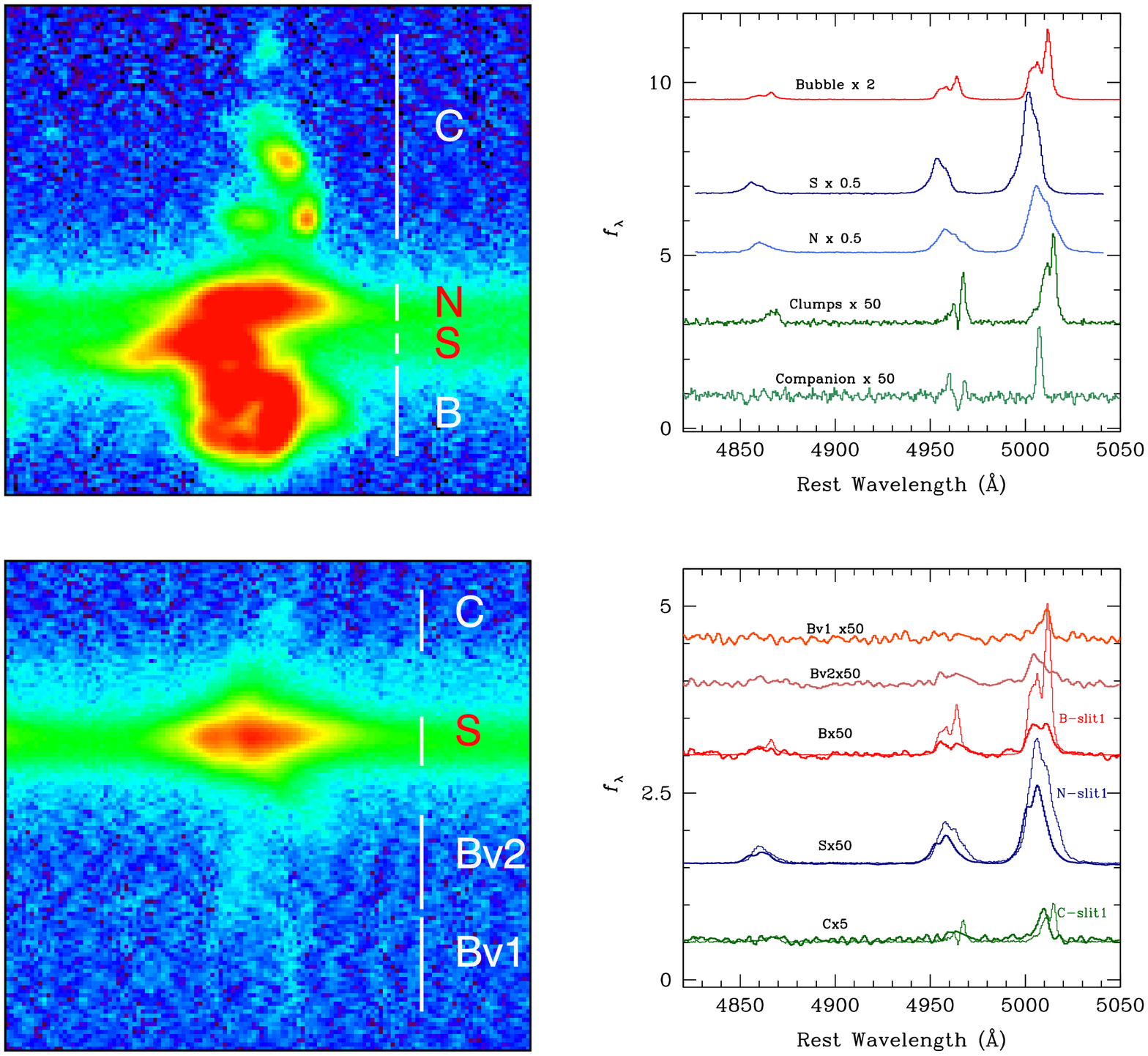,width=0.75\textwidth,keepaspectratio=true,angle=90}
}
\vskip -0mm
\figcaption[]{
  Two-dimensional spectra ({\it left}) and one-dimensional extractions
  along the labeled regions ({\it right}) for slit position PA1 ({\it
    top}) and PA2 ({\it bottom}).  The two-dimensional spectra are
  centered on 5007\AA\ and span 80\AA\ (rest-frame) in the spectral
  direction.  Each image is 23~\arcsec\ (50 kpc) in height, although
  the spectrum along PA2 extends 15 kpc further to the south to
  highlight Bv1 and Bv2 (the 'wisps').  In the two-dimensional spectra
  we label features of interest, including, from PA1 ({\it top left})
  -- {\bf B}: bubble; {\bf S}: southern nucleus; {\bf N}: northern
  nucleus; {\bf C}: clumps.  We also show the one-dimensional spectrum
  of the northern Companion seen in Fig. \ref{fig:bubbleim} with a 65 kpc
  projected distance from our program object.  We label in PA2 (bottom
  left) -- {\bf Bv1}: wisp feature 1, {\bf Bv2}: wisp feature 2, and
  {\bf C}: the small amount of flux in the same region as the clumps.
  The one-dimensional spectra ({\it right}) are extracted in the
  windows shown in the two-dimensional image, and the labels
  correspond, except that the {\bf B} spectrum {\it bottom right} is
  the sum of Bv1 and Bv2. The spectra have been multiplied by the
  factors indicated on the plot for ease of display. In addition, for
  the PA2 spectra ({\it bottom right}), we overplot the corresponding
  features from PA1 to demonstrate the relative velocities between the
  components. For instance, note that the bubble and the wisps Bv1 and
  Bv2 have a similar range of velocities. The artifact in the Clump
  and Companion spectra along PA1 are due to a gap between the two
  LDSS3 CCDs.
\label{fig:bubblelabel}}
\end{figure*}

In addition to stellar continua, there is strong, spatially resolved
\oiii$~\lambda\lambda 4959,5007$ line emission, as well as \hbeta\ and other
high-ionization forbidden lines. We focus here predominantly on the
\oiii.  A $\sim 51$ kpc-long \oiii\ emission line feature (denoted
`TT' for `tidal tail'; Fig. \ref{fig:bubbleim}) extends between \bub\
and the northern companion galaxy and somewhat beyond it. No associated stellar
continuum is detected for this feature, nor is it seen in the
broad-band image.  Additional likely tidal features are seen in the
broad-band image, such as the low surface brightness feature $\sim
10\arcsec$ (22 kpc) SE of the nuclei. This feature is tentatively
detected in \oiii\ emission in the two-dimensional spectrum
(see feature Bv2 in Figure \ref{fig:bubblelabel}).

\begin{deluxetable*}{cccll}
\tablenum{1}
\tablecolumns{5} 
\tabletypesize{\scriptsize}
\tablewidth{0pc}
\tablecaption{Spectral Energy Distribution \label{tableobs}}
\tablehead{ 
\colhead{$\nu$ (rest)} & \colhead{$\nu L_{\nu}$} & 
\colhead{$\sigma(\nu  L_{\nu})$} & \colhead{source} & \colhead{comment} \\
\colhead{(Hz)} & \colhead{($10^{44}$ erg~s$^{-1}$)} & \colhead{($10^{44}$ erg~s$^{-1}$)} & 
\colhead{} & \colhead{} 
}
\startdata
  2.22$\times 10^{15}$ & 0.934 & 0.097 & GALEX far-UV &
  http://galex.stsci.edu/GalexView/ \\
  1.49$\times 10^{15}$   & 0.817 & 0.049 & GALEX near-UV & same \\ 
  9.39$\times 10^{14}$   & 1.09 & 0.05 & SDSS $u$ &
  http://skyserver.sdss3.org/dr8/en/ \\
  6.93$\times 10^{14}$   & 1.77 & 0.08 & SDSS $g$ & \arcsec\ \\
  5.36$\times 10^{14}$   & 2.94 & 0.11 & SDSS $r$ & \arcsec\ \\
  4.37$\times 10^{14}$   & 4.61 & 1.01 & SDSS $i$ & \arcsec\ \\
  3.65$\times 10^{14}$   & 2.97 & 0.24 & SDSS $z$ & \arcsec\ \\
  2.70$\times 10^{14}$   & 3.35 & 0.03 & 2MASS $J$ & http://irsa.ipac.caltech.edu \\
  2.04$\times 10^{14}$   & 2.51 & 0.03 & 2MASS $H$ & \arcsec\ \\
  1.55$\times 10^{14}$   & 1.87 & 0.03 & 2MASS $K_S$ & \arcsec\ \\
  2.81$\times 10^{13}$   &$<$5.74 &    & IRAS 12$\micron$ & \arcsec\ \\
  1.35$\times 10^{13}$   & 10.1 & 0.6 & IRAS 25$\micron$ & \arcsec\ \\
  5.62$\times 10^{12}$   & 15.8 & 1.7 & IRAS 60$\micron$ & \arcsec\ \\ 
  3.37$\times 10^{12}$   & 7.2 & 1.7 & IRAS 100$\micron$ & \arcsec\ \\
  1.57$\times 10^{09}$   & 3.40$\times 10^{-4}$ & 0.09$\times 10^{-4}$
  & FIRST and NVSS & \citet{beckeretal1995,condonetal1998} \\
  4.10$\times 10^{08}$   &$<$3.62$\times 10^{-4}$&   & $<$0.25 Jy (80\%
  completeness limit) in Texas 365 MHz & \citet{douglasetal1996} \\
  8.31$\times 10^{07}$   & $<$8.22$\times 10^{-5}$&    & $<$0.28 Jy
  (2$\sigma$) at 74 MHz & \citet{cohenetal2007} \\
\enddata
\tablecomments{{\bf Spectral energy distribution of \bub\ from archival data.} 
SDSS fluxes are the average of Petrosian and cmodel
  fluxes. SDSS errors are half of the difference between Petrosian and
  cmodel fluxes combined in quadrature with the nominal error on the
  Petrosian fluxes. The large uncertainty in the $i$ band is
  presumably due to the highly disturbed morphology of the object, so
  that the standard SDSS model fits fail. FIRST and NVSS fluxes are 59.58 mJy and 62.9 mJy,
  correspondingly. The table entry lists the average of the two, with
  the uncertainty being half the difference between the two. Although
  the nominal threshold for the 74 MHz catalog is 5$\sigma=0.70$ Jy, there
  is no hint of any flux at the source position in the survey stamps,
  so a 2$\sigma$ upper limit is adopted. The far-infrared luminosity
of \bub\ as estimated using the fitting formula by \citet{sandersmirabel1996}
is $2.68\times 10^{45}$ erg/sec (the uncertainty is dominated by IRAS
flux calibration, $\sim 20\%$). We use an $h=0.7,
  \Omega_m=0.3, \Omega_{\Lambda}=0.7$ cosmology. \label{tab:sed}}
\end{deluxetable*}
\bigskip

The focus of this paper is on the well-organized kinematic feature to
the South of \bub, ``the bubble'' (see Figure \ref{fig:bubblelabel}).
The feature appears as a ring in the two-dimensional spectrum, with a
spatial extent of $\sim 12$ kpc and apparent velocity extent of $\sim
460$ \kms. The central velocity of the bubble at its base is close to that of
the N nucleus, $\sim 8$ kpc (projected) from its center. Equidistant
to the north of the N nucleus, we find three clumps spanning a comparable
velocity range. The bubble and the clumps are pure emission-line
features, with no stellar continuum detected in the spectra.  Emerging at a 
45\degr\ angle to the bubble are two 'wisps' of emission associated with the 
same outflow.

We subtract the stellar continuum and co-add the spectrum in velocity
space to evaluate the \oiii/\hbeta\ line ratio as a function of
position along the slit (Figure \ref{fig:ratio}). The ratio is
constant along the slit, with \oiii/\hbeta\ $\sim 10$, suggesting that
the gas is photoionized by an accreting BH.  This line ratio is
constant for a remarkable number of components, including both the S
and N nuclei, bubble, and the northern clumps. We have not corrected
these observations for extinction, but note that assuming $R_V = 3.1$
and the reddening law of \citet{cardellietal1989}, there is only an
$8\%$ difference in the extinction of \hbeta\ and \oiii.  Thus, if we
invoke (for instance) higher levels of extinction in the galaxy
center, we would still only expect the \oiii/\hbeta\ ratio to decrease
by $<10\%$ in the central region.

As argued in Paper II, the most plausible explanation for the constant
line ratios is that both nuclei host obscured quasars with similar
luminosities. Another possible scenario is that only one quasar is
present and it photo-ionizes the other nucleus; however, we find that
the fine-tuning required to satisfy the available observables makes
this hypothesis unlikely (Paper II). Future observations (in
X-rays with {\it Chandra} or in the radio with $\sim$1\arcsec\
resolution) will help distinguish between these two possibilities.

Recently \citet{fuetal2011ifu} presented integral field
observations of the 6\arcsec$\times$6\arcsec\ field around \bub\
(e.g., largely excluding the bubble itself). It is not obvious from
their \oiii\ map that they see emission line sources associated with
both nucleus N and nucleus S. They conclude that \bub\ harbors only
a single active black hole in nucleus N, and that the multiple kinematic
components are due to large-scale motions of the gas. It is possible 
that our slit did not intercept the peak emission from nucleus N,
thus explaining why the continuum strengths of nucleus S and nucleus
N are roughly comparable in our spectrum, even though source N is
considerably more luminous (Paper I). We still believe we are
seeing a distinct \oiii\ peak associated with nucleus S, but
cannot perform more detailed comparisons at this time.

Taking the spectrum of the bubble alone and subtracting the continuum,
we use the velocity profile of the \oiii\ line to fit all other
emission features in the spectrum and thus calculate line ratios
without modeling the complex velocity structure of the lines. In
particular, we find a ratio of
\oiii$~\lambda\lambda$4959,5007/\oiii$~\lambda$4363$=90.5\pm7.2$ and
therefore derive an electron temperature $T_e=13,500\pm 500$ K
(\citealt{osterbrockferland2006}. Furthermore, we detect the
[\ion{Ar}{4}]$\lambda\lambda$4711,4730 doublet with a ratio of $1.32\pm
0.37$, consistent with its low-density limit of $1.4$
\citep{kruegeretal1970}. Because the critical density for this 
transition is fairly high, even if we take our measurement as a lower
limit on the ratio, it places only a weak limit on the electron
density ($n_e\la 850$ cm$^{-3}$). The ratios of lines of the Balmer
series are consistent with no reddening.

The Magellan spectrum does not cover the density diagnostics
\oii$\lambda\lambda~3726,3729$ and
\sii$\lambda\lambda~6716,6731$, but both of these features are covered
by the SDSS spectrum taken in the 3\arcsec circular aperture centered
on the nuclei. The components of the \oii\ doublet are separated only
by 240 \kms\ and are far too blended to determine their ratio, given
that \fwoiii$=652$ \kms\ in this aperture. The \sii\ doublet has an
acceptable velocity separation, but the velocity structure of the
\oiii\ emission line does not provide a good fit to this feature, so
we cannot determine the line ratio in the same model-independent way
as for the Magellan spectrum. We measure the peak ratio of the
two components of the doublet to be 1.20, which is likely to be a
lower limit on the actual ratio because of blending. Taking
\sii$~6716/6731 >$1.2 yields $n_e<250$ cm$^{-3}$.

The object is strongly linearly polarized, with the polarization fraction
reaching 8\% at shortest observed wavelengths (Figure
\ref{fig:polarim}). These observations are consistent with the
hypothesis that \bub\ is powered by a luminous quasar whose emission
is largely blocked along the line of sight to the observer, but
escapes along some other directions, and then scatters off the
interstellar material within the host galaxy
\citep{smithetal2003,zakamskaetal2005} and reaches the observer. In
particular, the emission from the quasar escapes predominantly in the
N-S direction in the plane of the sky, as indicated by the 
$\sim$90$^{\circ}$ PA of the plane of polarization.
Furthermore, the wavelength
dependence of the polarization fraction can be attributed to the
dilution of the scattered quasar light by the unpolarized emission
from the extended narrow-line region and starlight from the host
galaxy. Indeed, the dilution of the polarized flux by starlight
increases to the red, as expected from a red stellar population, and
an abrupt minimum in P is seen around \oiii\ (Figure \ref{fig:polarim}). The
generally higher values of P for the N-S slit orientation of the May
27, 2011 observation are likely caused by the fact that the spectral
extraction aperture runs along the axis of the presumed scattering
cone, thereby including a larger fraction of the scattering regions
than the earlier measurement made with the Kuiper Telescope. 
Other small discrepancies between the two
spectropolarimetric observations are likely caused by differing
positions of the object within the slit and by the difference in the
slit direction, suggesting that polarized structures within the source
are partially resolved. 

\subsection{Observed Kinematics and Geometry}

We propose that \bub\ is powering a large-scale outflow, which is
entraining the warm ionized gas that we see radiating in \oiii.  Here
we present the observed kinematics along each slit position that we
will use to estimate the energetics of the outflow.

Starting with the bubble along the first (NS) slit position PA1, we
take two-pixel--wide extractions along the spatial direction starting
from nucleus S and working our way to the bottom of the bubble (all
within the bubble as demarcated in Figure \ref{fig:bubblelabel}). We
use Gaussian fits to identify the primary velocity components at each
spatial position. These velocities are plotted in Figure
\ref{fig:bubblemodel}, ignoring some weak components. This
velocity pattern, increasing symmetrically to a projected velocity of $\sim
250$~\kms, is the signature of an expanding 
sphere of gas with limb brightening. 

In addition to the bubble, there are three clumps to the north of
nucleus N that we see in slit position PA1. We note that (a) their
central velocity is similar to that of the bubble and (b) the maximum
velocity extent across the clumps is comparable to that seen in the
bubble (see Figure \ref{fig:bubblelabel}). Therefore, we suggest that
these clumps are the northern side of the same outflow.  The bubble is
systematically blueshifted by $\sim215$~\kms\ relative to the clumps;
therefore, the axis of symmetry of the outflow is likely not quite in
the plane of the sky but is tipped so that the bubble is expanding
slightly towards us and the clumps are from the component moving
slightly away from us (see Figure \ref{fig:bubblemodel}). The northern
and southern components of the outflow are discontinuous; we suggest three
possible corresponding geometries. One possibility is that we are
seeing two unrelated components, but given the similarity of the
systemic velocities and the velocity spread we find this
unlikely. Another possibility is that our slit did not pass through
the center of the outflow, but instead was 1$-$2\arcsec\ (2.3-4.6 kpc)
offset (see arguments above suggesting that the PA1 slit was not
centered on nucleus N). In this case an outflow that forms a figure
`8', with the center taken out, could appear as a pair of disconnected
features. Yet another possibility (that we favor on the basis of
kinematic model presented in the next section) is that the merger
provides plentiful gas in the inner several kps of the system,
pinching the flow, which finally breaks free of this dense gas on
either side, as shown in Figure 5.

The signal-to-noise ratio along PA2 is not very high, and thus we do not
measure the velocities as a function of position in this case.
The two linear structures to the south of
nucleus N, Bv1 and Bv2 (Figure \ref{fig:bubblelabel}) show a 
clear trend of increasing line-of-sight velocity with projected
distance. These wisps represent gas that is escaping at a
45\degr\ angle to the gas in the bubble.

\subsection{Kinematic Model}

We see outflowing gas to the north, south, and south-east of the
nuclei of \bub.  The emission-line gas is clearly expanding away from
the nuclei.  Is it being actively driven from the nuclei and
accelerating away, or is it decelerating as it slams into the
interstellar or intergalactic medium?  The linearly expanding
structures along PA2 (wisps Bv1 and Bv2) are the tell-tale signs of an
accelerating outflow \citep{veilleuxetal1994}. Both a conical outflow
with $v_{3D}\propto r_{3D}$ and a spherical outflow blown from one side with
$v_{3D}\propto r_{3D}^2$ will produce $v\propto r$ when projected onto
the line of sight ($v$) and on the plane of the sky ($r$). The cap of
the bubble along the southern-most edge is produced by a shell where
the outflow interacts with gas outside the galaxy, either tidal debris 
or the intergalactic medium. Because of
projection effects, the cap is brightest in the plane of the sky,
where its line-of-sight velocity is zero, so we do not know whether
the shell is still accelerating or whether it has already started to decelerate.

If we assume that there is a single accelerating agent causing all of the
outflowing components, then a credible picture of the geometry can be built. Due to
interactions with the ambient medium of varying density, the outflow
is complicated enough that we cannot uniquely determine its geometry
from two slit positions, but we are able to bracket the outflow
velocities using a reasonable range of models. Our primary goal is to
explain the symmetric velocity structure along the bubble in a way
that is also consistent with the clumps to the north and the linearly
increasing velocity structures ('wisps') to the south-east along PA2.

We first consider an accelerating ($v_{3D} \propto r_{3D}$) biconical outflow,
which is a natural geometry for an obscured quasar and easily explains
the wisps. 

\begin{figure*}
\vbox{ 
\vskip 0.2truein
\hskip +0.2in
\psfig{file=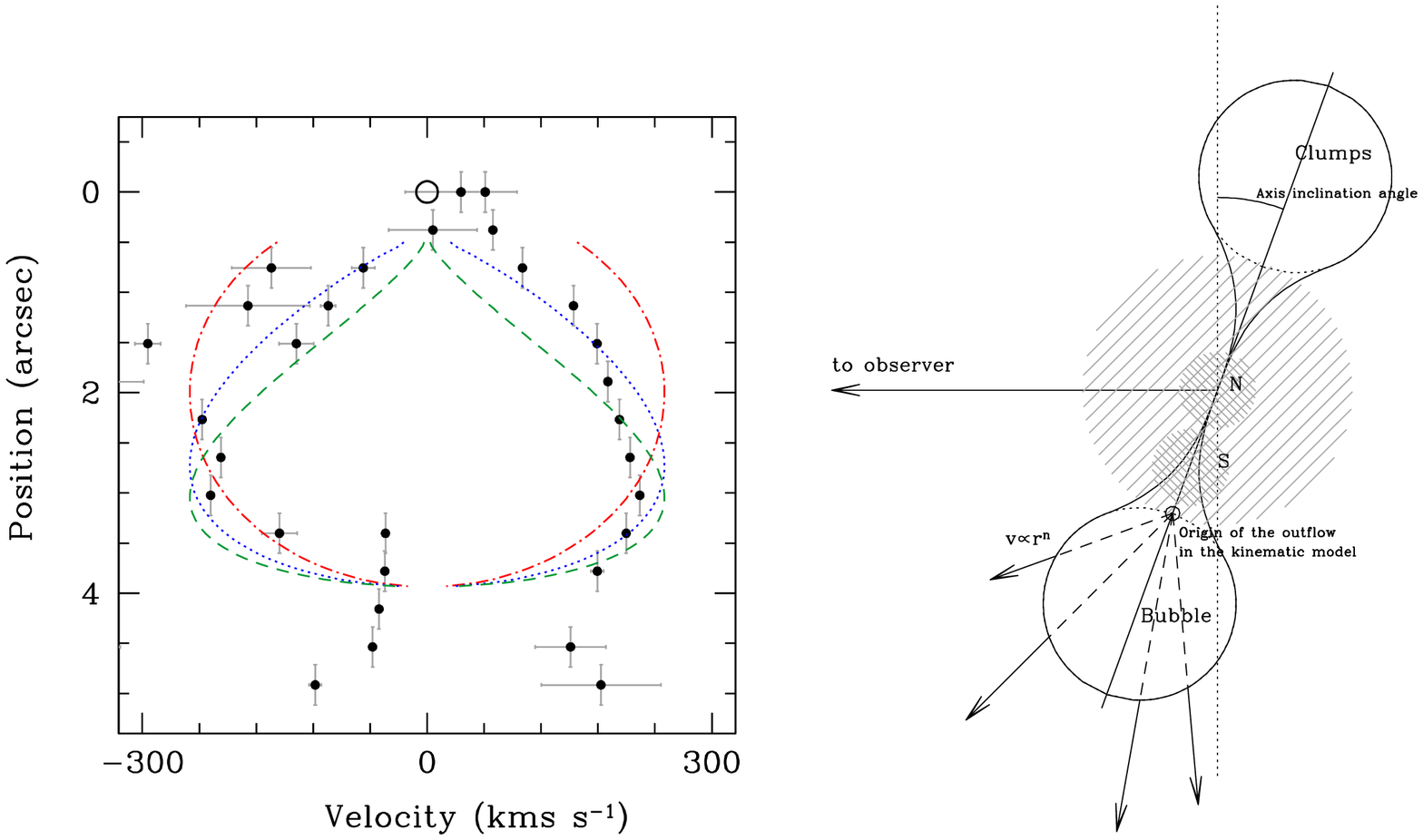,width=0.5\textwidth,keepaspectratio=true,angle=-90}
}
\vskip -0mm
\figcaption[]{
We show our simple kinematic model ({\it left}) and an accompanying cartoon 
of the outflow region ({\it right}).
{\it Left}: Position-velocity diagram centered on the the \oiii\
bubble, as demarcated in Figure \ref{fig:bubblelabel}. Filled black
circles indicate velocity peaks measured at intervals of 0\farcs4.  We
assume that the accelerating agent is offset from nucleus N by $\sim
2$\arcsec, thus the outflow is detached from the continuum source.
While the acceleration arises from nucleus N, we show only the
outflowing region, starting at the open circle at the origin.
Overplotted is a schematic model of a quasi-spherical outflow that is
decelerated at its base as it passed through continuum sources N and
S.  The velocity along the flow increases with radius from source N
with a power-law dependence $\upsilon \propto r^n$. We plot models
with $n=1$ ({\it red dash-dot}), $n=1.5$ ({\it blue dotted}) and $n=2$
({\it green dashed}).  This model yields a maximum outflow velocity of
$\sim 1000$ \kms.
{\it Right}: A schematic cartoon illustrating our best guess of the system geometry, 
including the two continuum sources, the Bubble to the south, Clumps to the 
north, and the slight tilt in the plane of the sky relative to the observer.  The circle 
indicated the origin of our model to the left.
\label{fig:bubblemodel}}
\end{figure*}

\vbox{ 
\vskip +45mm
\hskip 0.1in
\psfig{file=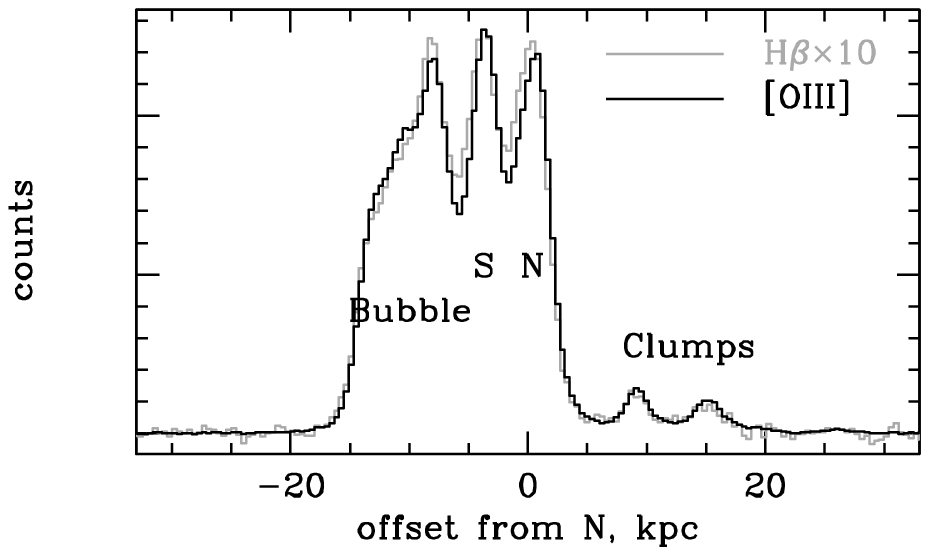,width=0.45\textwidth,keepaspectratio=true}
}
\vskip -0mm
\figcaption[]{
The spatial profiles of [O {\tiny III}] and \hbeta\ emission along the N-S slit. 
The [O {\tiny III}]/\hbeta\ ratio is constant and close to 10.
\label{fig:ratio}}
\vskip 5mm
\noindent
This model fails to reproduce the concave velocity
structure (the southern ``cap'') observed in the bubble (Figure
\ref{fig:bubblemodel}).  Another straightforward model is a 
spherical outflow centered on nucleus N. If only the shell is emitting,
it is then difficult to reproduce $v \propto r$ at small
radii in the bubble and the wisps. Both the cap and the
accelerating features are naturally reproduced in the framework of a
bubble blown from its side, not from its center, perhaps due to the
high density of matter near the energy injection point
\citep{veilleuxetal1994}. This model is reminiscent of outflows in
starburst galaxies, in which propagation in the plane of the galaxy is
strongly suppressed and the outflows escape predominantly
perpendicular to the plane \citep{veilleuxetal2005}. Note that if the
outflow axis is inclined toward the plane of the sky, then the
velocity offset between the clumps and the bubble (relative to the
maximum span in the velocity of the bubble) leads to an estimate of
the inclination of the axis of $\sim 20$\degr, which could in
principle boost the space velocities relative to those assumed here.

A simple version of this model is shown in Figure
\ref{fig:bubblemodel}. We assume that the flow is spherical, with the
origin on the side of the sphere and the axis in the plane of the sky,
that only the shell is emitting and that $v_{3D} \propto r_{3D}^n$
where $n = 1,1.5,2$. Then we project the system onto the sky and the
velocities along the line of sight and plot the model curves along
with the observations of the bubble in \bub. We perform a \chisq\ fit to the 
observed position-velocity diagram taking $1 < n < 2.5$. 
We exclude all data points with velocities greater than
$\pm 400$~\kms, since they are not treated by our model.  We further
use a minimum velocity error of $20$~\kms\ (a fraction of 
a pixel).  Models with $1.2 < n < 1.5$ all yield a reduced 
\chisq\ differing from the minimum by less than one.  The corresponding range of 
projected velocities is $250 < v_{3D} < 1000$~\kms.  
At the lowest velocity there is no projection effect: the maximum observed 
velocity is also the maximum space velocity.  This is a strict lower 
limit to the energetics.

We postulate that all of the line emission is part of a
quasi-spherical outflow blown by the AGN, that extends from the
southern-most edge of the bubble to the clumps in the north
(illustrated schematically in Figure \ref{fig:bubblemodel}). However,
due to interference from gas and dust in the central pair of galaxies,
the outflow cannot escape freely in all directions.  Along PA2, the
wisps represent gas that has broken free and is in free expansion.  In
contrast, along the N-S direction (PA1) the outflow runs into the two
galaxies.  The ring shape in the PA1 2D spectrum results from gas
breaking out of the high density material associated with galaxy S and
accelerating away from the merger with $v_{3D}\propto r_{3D}^{1.5}$.
In addition, the structure is disconnected from its origin (nucleus N)
because of the intervening matter. The clumps in PA1 are the northern
counterpart of the same flow, produced closer to nucleus N than the
bubble since there is less material to penetrate. In this picture, the
similar velocity spread seen along the bubble, clumps and wisps
(Fig. \ref{fig:bubblelabel}) are naturally explained as arising from
the same outflow. The opening angle of the outflow is no less than the
45\degr\ angle between PA1 and PA2, since both slit orientations
captured it.

\subsection{Energetics}

With our crude outflow model, and thus deprojected velocities, we
can now estimate the energy injection $\dot{E}$ required to power the
observed bubble, following \citet{nesvadbaetal2006}. 
One method is to estimate the kinetic energy in the shell, based on
an estimate for the gas mass and the observed velocity, and then
divide by the dynamical time to estimate an energy injection rate
\citep[e.g.,][]{martin1998}. The other approach is to treat the bubble like a
supernova remnant, and estimate the energy needed to expand the bubble
into a low-density ambient medium.  In both cases we are forced to
make assumptions, but with the two methods we can at least bracket the
kinetic luminosity of the outflow. 

First, we derive a lower limit on the energy in the outflow based on
its apparent kinetic energy. The dominant uncertainty here is the mass
of gas in the bubble.  In principle, with an electron density, we can
turn the observed \hbeta\ luminosity ($L_{\hbeta}=7.5 \times
10^{41}$~ergs~s$^{-1}$) into a hydrogen gas mass. As explained in
detail in Paper II, we have two methods to estimate gas density in the
extended nebulae of obscured quasars.  One
is the standard diagnostic line ratios, which give density estimates
$n_e \la 1000$ cm$^{-3}$. Unfortunately, this method is biased toward
regions of the brightest line emission, and thus regions of the
highest density.  Furthermore, in clumpy media, density diagnostics
often yield values close to the critical density in a wide range of conditions
(Paper II). The other method is to use the observations of scattered
light at large distances from the nucleus to infer a gas density, but this
has only been possible for a couple of objects observed with the {\it
  HST} and with particularly well-measured scattered light nebulae
\citep{zakamskaetal2006}.  This method gives a mean electron density
of $n_e \la 1$ cm$^{-3}$. Clearly, the two estimates are 
discrepant, suggesting that gas clumping is significant. Defining the
clumping factor as $\kappa=\langle n_e^2\rangle/\langle n_e\rangle^2$,
we can write the mass in gas as:
\begin{equation}
M_g = 1.7 \times 10^9 M_{\odot} \cdot L^{41}_{\rm H\beta} \langle n_e\rangle^{-1} \kappa^{-1}.
\end{equation}
In section \ref{sec:obsem}, we used diagnostic line ratios to find
that in the nuclear region $n_e<250$ cm$^{-3}$ and that \oiii/\hbeta\
is constant throughout the nuclei and the bubble. Since at the
location of the bubble the ionizing source is 5-10 kpc away, the
density must drop with radius by at least a factor of 20$-$100 to
maintain the same ionization parameter (the ratio of the number of
ionizing photons to electrons, $U$).  This decrease can plausibly be
somewhat offset by a clumping factor of similar magnitude.  Therefore,
our final mass estimate is $M_g\ga 5\times10^7$~\msun. This mass is a
lower limit, because it takes into account only the ionized gas that
produces emission lines. Any lower density gas at higher temperature
filling the bubble would not be included in this estimate.  

\vbox{ 
\vskip 38mm
\hskip 0.1in
\psfig{file=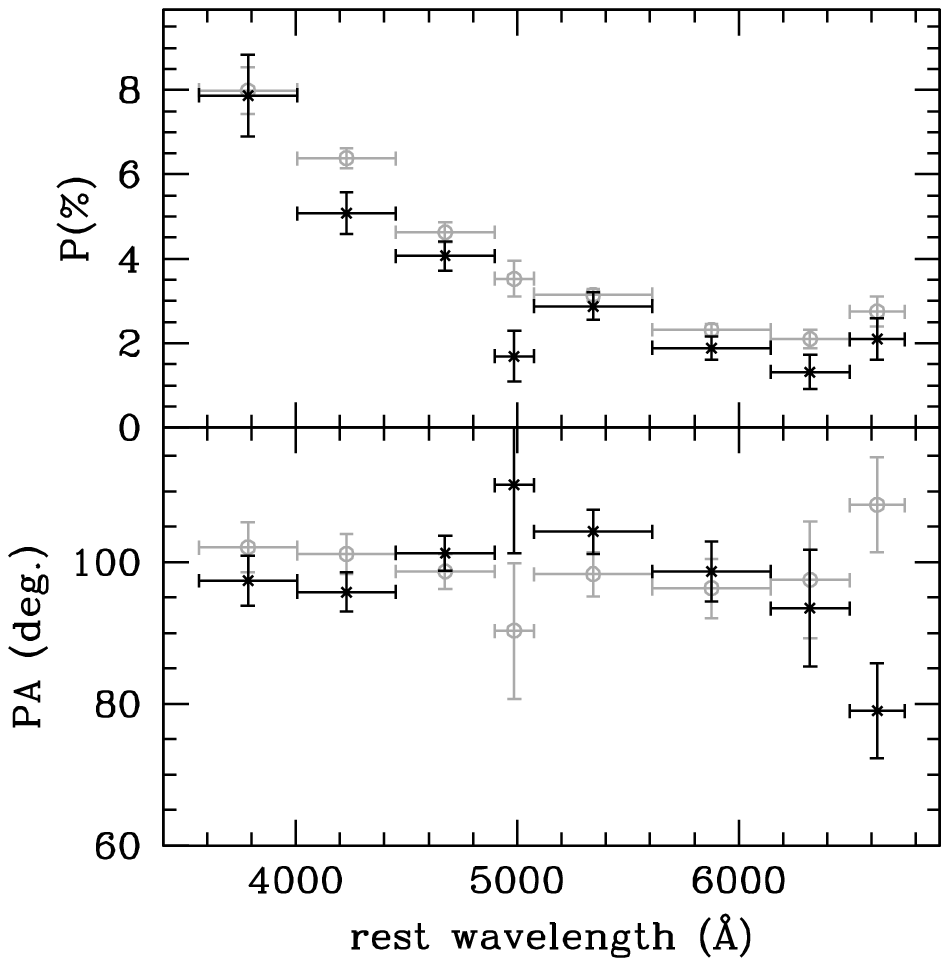,width=0.45\textwidth,keepaspectratio=true}
}
\vskip -0mm
\figcaption[]{
{\it Top}: Polarization fraction. {\it Bottom}: polarization position
angle (degrees E of N) as a function of rest-frame wavelength. Black
skeletal points are from the Kuiper telescope observations, with the
3\arcsec-wide slit in the E-W direction, and and grey open circles  
denote binned spectropolarimetry from the Bok telescope, with 
the 3\arcsec-wide slit oriented N-S.
\label{fig:polarim}}
\vskip 5mm
\noindent
On the other hand, without a measurement of the clumping factor $\kappa$, we
cannot say for sure what fraction of the observed volume is filled by
high-density gas.  In the case of the observed bubble in NGC 3079
\citep[e.g.,][]{filippenkosargent1992,veilleuxetal1994}, the
energetics estimates were revised downwards based on high-resolution
\hst\ imaging of the ionized gas filaments \citep{ceciletal2001}.  We
plan to obtain integral-field observations of \bub\ in the Spring,
which will hopefully help settle the outstanding ambiguities.

Based on the extent of the bubble, $R=12$~kpc, and the maximum
deprojected velocity of $v=250-1000$~\kms, we find the dynamical time, $t_d \approx
R/v \approx 10-40$ Myr. It is interesting to note that this
timescale is similar to typical AGN lifetimes
\citep{martiniweinberg2001,martini2004}. With our previous estimate of
the mass of the emitting gas, this corresponds to a mass outflow rate of
$\dot{M} \sim M_g/t_d > 2-5$ \msun~yr$^{-1}$. The corresponding kinetic
energy is $E_{\rm kin} \sim M_g v^2/2 > 3-50 \times 10^{55}$~ergs.  For a
constant energy injection rate over the lifetime of the bubble, we
find $\dot{E} > 3-10 \times 10^{41}$~ergs~s$^{-1}$. This is the kinetic energy of
the emission line gas alone and, therefore, a strict lower limit on the total
energy of the outflow. 

The energy injection rate required to expand the bubble into a
low-density medium, $n_0$, in the galaxy halo can be calculated in a
manner similar to the Sedov-Taylor solution for supernovae.  Following
\citet{nesvadbaetal2006}, we assume that the radiative losses are
minimal and that energy injection rate is constant rather than a
single explosion. We then find
\begin{equation}
\dot{E} \approx 1.5 \times (\Omega/4\pi) 10^{46} R_{10}^2 v_{1000}^3 n_0~{\rm ergs~s^{-1}},
\end{equation}
where the extent of the bubble $R_{10}$ is in units of 10 kpc,
$v_{1000}$ is the expansion velocity in units of $1000$~\kms and
$\Omega/4\pi$ is the covering factor of the outflow in
steradians. Taking $R_{10}=1.2$, $n_0=0.5$~cm$^{-3}$, and $v_{1000} =
1$, $\dot{E} \approx (\Omega/4\pi) \times 1.1 \times
10^{46}$~erg~s$^{-1}$.  If we instead adopt the lower limit on
velocity (i.e., no deprojection) of $v_{1000} = 0.25$, we find
$\dot{E} \approx (\Omega/4\pi) \times 2 \times 10^{44}$~erg~s$^{-1}$.

In addition to the deprojected velocity, the major uncertainty in this
value comes from the lack of good measurements of $n_0$ and
$\Omega$. Both slit positions, separated by 45\degr, captured the
signature of the flow and \citet{fuetal2011ao} find a strong component
of \oiii\ emission to the northeast of N, which is not covered by our
observations.  We therefore conclude that $\Omega$ is not too small.
It is much harder to estimate the proper value for $n_0$.  The \hst\
observations of \citet{zakamskaetal2008} alluded to above find clear
evidence for electron scattering off of a medium with $n_e \la 1$ cm$^{-3}$, 
even at these large projected distances of $\sim 10$ kpc from the nucleus. 
We clearly see evidence for denser gas from the observed line emission. 
Thus, we view our choice of $n_0$ to be a conservative lower limit.

The kinetic energy injection rate of the outflow is bracketed by the
values $10^{42}$~erg~s$^{-1}$ (where we likely underestimated the
amount of mass by a large factor) and $10^{46}$~erg~s$^{-1}$ (where we likely
overestimated the outflow covering factor and the density of the
ambient medium by smaller factors). We will
therefore adopt a fiducial range of $10^{44-45}$~erg~s$^{-1}$ for the
kinetic energy injection rate in the discussion that follows. It is
interesting to note that our upper limit on the kinetic energy
injection rate is roughly consistent with our estimate for the
radiative bolometric luminosity of $L_{\rm bol} \sim
10^{46}$~ergs~s$^{-1}$. 

\subsection{Can the outflow be driven by star formation?}

Can the energy of the outflow be plausibly supplied by any source
other than accretion onto the supermassive black hole?  Vigorous star
formation can lead to large-scale outflows \citep{veilleuxetal2005}
that are observable both in ionized gas in emission
\citep[e.g.,][]{chevalierclegg1985,heckmanetal1990} and in neutral gas
in absorption \citep[e.g.,][]{rupkeetal2005big,martin2005}. In an
active merger like \bub, with plenty of gas available (as evidenced by
our emission line observations), it is natural to expect that some
on-going star formation is present. While from the high-ionization
line ratios (such as the uniformly high \oiii/H$\beta\simeq 10$) it is
clear that the photo-ionization of the gas is completely dominated by
the quasar, star formation may still be the primary driver of the
kinetic outflow. Unfortunately, despite the availability of abundant
multi-wavelength data and deep spectra of the host galaxy, the
relative contributions of star formation and obscured quasar activity
to the bolometric output of \bub\ are very hard to determine. This is
a generic difficulty arising due to the fact that most of the flux
from the object is coming out at mid- and far-infrared wavelengths,
where the emission is due to thermal radiation of dust particles
heated by young stars or the quasar in a manner notoriously
insensitive to the spectrum of the underlying power source.

The bolometric luminosity of the quasar, as expected from its \oiii\
luminosity \citep{liuetal2009}, is $1.5\times 10^{46}$ erg~s$^{-1}$,
although such scalings have dispersion of $\ga 0.5$ dex.  The
bolometric luminosity estimated using the 25 and 60\micron\ IRAS
fluxes and quasar bolometric corrections \citep{richardsetal2006}
ranges between $7.2\times 10^{45} - 2.4\times 10^{46}$
erg~s$^{-1}$. The reasonable agreement between these values indicates
that most, if not all, of the mid- and far-infrared luminosity from
\bub\ ($2.7\times 10^{45}$ erg~s$^{-1}$) can plausibly be produced by
the quasar activity rather than star formation.  The high degree of
polarization likewise suggests that the AGN is dominant.

If, however, all of the infrared luminosity of this source were powered by star
formation, it could just barely power the observed outflow.  We use the
calculation of the mechanical energy injection rate from a constant
rate of star formation by \citet{leithererheckman1995} and
\citet{veilleuxetal2005}. For every solar mass of stars formed, the
combination of stellar winds and supernovae can produce at most
$L_{\rm mech} \approx 7\times 10^{41}$~ergs~s$^{-1}$ at time $t
\approx 10^7$~yr. Thus, to produce a $10^{44-45}$ erg~s$^{-1}$ outflow
would require $>100-1000$~\msun~yr$^{-1}$ of star formation and
predict a star-formation luminosity of $L_{\rm FIR}=3\times
10^{45-46}$~ergs~s$^{-1}$.  Thus, star formation is marginally
consistent with powering the outflow at the lower limit of the energy
range.  On the other hand, a starburst of $>100$~\msun~yr$^{-1}$ is
hard to reconcile with our optical spectroscopic and spectropolarimetric 
observations.  We see
no evidence in the spectrum for an intermediate age population (i.e.,
the Balmer absorption lines are consistent with an old stellar
population), and the blue continuum that we do see is consistent with
scattered light \citep[e.g.,][]{liuetal2009}.  Therefore, we consider
power from star formation as unlikely given the currently available
data.

\vskip 5mm

\subsection{Can the outflow be jet-driven?}

Radio flux densities at 1.4 GHz can be obtained from the Faint Images
of the Radio Sky at Twenty cm survey (FIRST; \citealt{beckeretal1995})
and the NRAO VLA Sky Survey (NVSS; \citealt{condonetal1998}). The
FIRST detection is unresolved -- the nominal deconvolved size of
1.2$\times$ 0.6\arcsec\ is smaller than half of the beam size of
5.4\arcsec\ \citep{houlvestad2001}.  The FIRST peak flux is 57.90
mJy~beam$^{-1}$ and the integrated flux is 59.58 mJy; the ratio of
these two measures is consistent with that seen in point sources
\citep{ivezicetal2002, kimballivezic2008}. The high-resolution FIRST
observations tend to resolve out the low surface-brightness extended
emission, so the comparison with lower-resolution NVSS observations
(45\arcsec\ beam) provides yet another measure of extended
emission. The NVSS catalog flux is 62.9 mJy, and again the difference
between NVSS and FIRST-measured fluxes is consistent with that seen in
point sources \citep{kimballivezic2008}. From these observations, it
appears that \bub\ lacks extended radio emission on $>4$ kpc
(2\arcsec) scales at the few mJy level.

Using the radio-infrared correlation of star-forming galaxies
\citep{helouetal1985} and its convenient normalization presented by
\citet{moricetal2010}, we find that no more than 20\% of the observed
radio emission can be due to star formation. Thus, most of the radio
emission is attributable to one or both quasar nuclei. Nevertheless,
the position of the \bub\ in the \oiii-radio diagram
\citep{xuetal1999,zakamskaetal2004,lalho2010} puts it squarely into
the radio-quiet regime, 1.5$-$3 dex below radio-loud sources in
\citet{nesvadbaetal2008} and \citet{fustockton2009}.  For direct
comparison with \citet{xuetal1999}, we use a radio spectral index of
$\alpha=-0.47$ (defined as $F_{\nu} \propto \nu^{\alpha}$, see below)
to convert a rest-frame-corrected value of $\log\,(P_{\rm
  1.4GHz},\,{\rm erg~s^{-1}~Hz^{-1}})= 31.32$ into $\log\,(P_{\rm
  5GHz},\,{\rm erg~s^{-1}~Hz^{-1}})=31.06$ at $\log\,(\loiii,\,{\rm
  erg~s^{-1}})=42.78$ taken from \citet{reyesetal2008}. Obscured
quasars with such radio luminosities have only a $\sim 10\%$ chance to
have radio extents $>3$ kpc \citep{lalho2010}, consistent with the
lack of any sign of extended emission in \bub.  In principle, we
cannot rule out a $<4$ kpc-scale jet as the accelerating agent, as has
been proposed for NGC 3079 \citep{ceciletal2001}.  On the other hand,
the outflow we observe is considerably more extended and more
energetic than that in NGC 3079, so it is not clear that energy
injection on ~1 kpc scales can power the observed 20 kpc outflow.
Deeper radio interferometry at 1\arcsec\ resolution would settle the
issue.

The source is undetected in the Texas survey at 365 MHz
\citep{douglasetal1996} and in the VLA Low-Frequency Survey at 74 MHz
\citep{cohenetal2007}, with the 74 MHz observation providing a
stronger upper limit, so we focus on this observation. This implies a
limit on the spectral index $\alpha>-0.47$ (hence the use of this
value in the calculation above). Such a flat spectrum effectively
excludes a ``relic'' radio source \citep{komissarovgubanov1994}.

High-resolution, high-sensitivity radio observations would be useful in
establishing the presence of two quasars and of any extended radio
emission, especially on $5-10$\arcsec\ scales. Since the separation of
the nuclei is $\sim 1$\arcsec, observations with VLA would be well-suited
for these purposes. The archival radio observations
available to us at the moment indicate that the outflow in \bub\ is
unlikely to be jet-driven.

\section{Conclusions}

\bub\ is an on-going galaxy merger at redshift $z=0.123$ that 
contains at least one obscured quasar in its northern nucleus. The
bolometric luminosity of the object, estimated at $1.5\times 10^{46}$
erg~s$^{-1}$, places it among the more luminous nearby quasars 
\citep[e.g.,][]{hopkinsetal2007}. We have conducted a study of the
kinematics and spatial distribution of the ionized gas in this
source. We find a $\sim$ 50 kpc tidal feature close to the systemic
velocity of the object that is photo-ionized by the quasar along an
unobscured direction. High optical linear polarization is observed, also 
indicating that the emission-line regions are powered by the AGN.
In addition, we observe a 
pair of expanding shells, or bubbles, centered on the northern nucleus
and expanding into the intergalactic medium. The remarkable scale of the
expanding structure ($\sim 40$ kpc in total extent), its high expansion
velocities (observed velocity range of 460 \kms, with physical velocities likely 
reaching 1000 \kms), and its large opening angle and
luminosity imply that the kinetic energy of the outflow likely exceeds
$>10^{44-45}$ erg~s$^{-1}$. 

We find that while star formation may be occurring concurrently with
quasar activity in \bub, it is energetically insufficient to power the
observed outflow, likely by an order of magnitude, as further
supported by the polarization observations. The merging quasars are
radio quiet, and there is no evidence for extended radio emission;
therefore, the outflow is unlikely to be powered by a relativistic jet
in a manner similar to outflows seen in some radio galaxies. We
propose that \bub\ may be an example of on-going feedback
from a quasar in its most common radiative phase. Such feedback has
been postulated to account for the colors and luminosities of massive
galaxies \citep[e.g.,][]{springeletal2005}, but it has been difficult
to assemble direct evidence of interactions between radio-quiet
quasars and their large-scale gaseous environments aside from
individual cases noted in the introduction. \bub\ provides compelling
evidence that accretion energy can drive wide-angle outflows of gas
beyond the confines of the host galaxy.

The presence of extended ionized emission by itself is not a signature
of quasar feedback \citep{stocktonmackenty1987}. If one could obtain
arbitrarily sensitive 
observations, one would be bound to find small amounts of gas, and if
this gas finds itself along an unobscured direction to the quasar it
can be photo-ionized out to large distances. For example,
\citet{villarmartinetal2010} find a 180 kpc emission-line feature near
an SDSS type 2 quasar, but its spatial distribution and kinematics
make it clear that the feature is tidal debris illuminated by the
quasar. Images of this source taken with the \emph{Hubble Space Telescope}
\citep{zakamskaetal2006} demonstrate that the scattering regions
(and therefore illumination cones) have the same E-W orientation as
the 180 kpc tidal tail. We find a very similar feature in \bub\
extending in the N-S direction, which also happens to be the direction
of quasar illumination, as indicated by the polarimetric
observations. Rather than swept from the galaxy by quasar feedback to
50-100 kpc away, the gas seen in these two cases is most likely tidal
debris  which happened to be along an unobscured line of
sight to the quasar. 

Similarly, \citet{liuetal2009} find several
obscured quasars with companions out to $\ga 30$ kpc away from the
quasar showing high equivalent width emission lines with
high-ionization line ratios. Again, rather than instances of quasar
feedback, these occurrences are probably small companion galaxies
which happen to fall within one of the quasar illumination cones. 

What sort of signatures specifically implicate quasar feedback? The
absence of stellar continuum is a necessary (although not a
sufficient) requirement, as it helps rule out illuminated companion
galaxies and tidal features. Gas velocities that exceed the escape
velocity of the galaxy are a clear sign, but since the outflows from
type 2 quasars are directed in the plane of the sky, the chances of
observing such high velocities in these objects are low.  The presence
of a kinematically organized and symmetric structure, such as that
seen in \bub, is a very compelling clue, as it completely rules out an
instance of a gas inflow illuminated by the quasar (while the nature
of radial velocity observations technically makes inflow and outflow
scenarios degenerate, it is hard to imagine the gas conspiring to
accrete in such symmetric manner).  The high linear-polarization
fraction of $P=8\%$ provides unambiguous evidence that there is a
luminous obscured quasar in \bub\ and severely limits the fraction of
the bolometric luminosity of this source that may be due to star
formation. Furthermore, through the measurement of the polarization
position angle, polarimetric observations indicate that the outflow
direction (N-S) is well aligned with the predominant direction of
quasar illumination.  Clearly a narrow-band image or 
integral-field spectroscopy would provide a more complete picture of 
the energetics.

We did not find similarly well-organized kinematics in the rest of the
Magellan sample \citep{greeneetal2011}, but the large velocity widths
and the lack of rotation seen even in prominent disk galaxies observed
along their semi-major axes make quasar feedback the likely
explanation in those cases as well.  Our work to study three-dimensional 
outflows in luminous obscured quasars is ongoing with integral-field 
spectroscopy, which we hope will provide better statistics on the population 
as a whole (Zakamska, N. L. et al. in preparation).

\acknowledgements
The referee made valuable comments that significantly improved this
manuscript.  We gratefully acknowledge useful conversations with
Daniel Proga.  We also thank L.~C.~Ho and A.~J.~Barth for invaluable
assistance in the earlier stages of this project.  P.S.S. acknowledges
support from NASA/Fermi Guest Investigator Program grant NNX09AU10G.

\end{document}